\documentclass[12pt]{article}
\usepackage{jheppub}
\newcommand\I{{\mathrm i}}
\newcommand\de{{\mathrm{d}}}
\def\CA{{\cal A}}
\def\CB{{\cal B}}
\def\CL{{\cal L}}
\def\CM{{\cal M}}
\def\CN{{\cal N}}
\def\CS{{\cal S}}

\title{Opers and TBA}

\author[1]{Davide Gaiotto}

\affiliation[1]{Perimeter Institute for Theoretical Physics\\%
31 Caroline Street North, ON N2L 2Y5, Canada}

\abstract{In this note we study the ``conformal limit'' of the TBA equations which describe the geometry of the moduli space of four-dimensional ${\cal N}=2$ gauge theories 
compactified on a circle. We argue that the resulting conformal TBA equations describe a generalization of the oper submanifold in the space of complex flat connections on a Riemann surface. 
In particular, the conformal TBA equations for theories in the $A_1$ class produce solutions of the Schr\"odinger equation with a rational potential.}

\begin{document}
\maketitle
\section{Introduction}
The moduli spaces $\CM$ of $\CN=2$ four-dimensional gauge theories compactified on a circle
are a rich subject of investigation. They are endowed with an hyperk\"ahler metric, 
which encodes the BPS spectrum of the four-dimensional theory. For theories of the class $\CS$, 
which arise from the compactification of 6d SCFTs on punctured Riemann surfaces, 
the moduli spaces $\CM$ coincides roughly with the moduli spaces of solutions of Hitchin's equations,
which play an important role in mathematical physics and mathematics. 

The connection between the moduli spaces and the BPS spectrum was used in \cite{Gaiotto:2008cd} 
to set up a system of integral equations which compute the  hyperk\"ahler metric
at any given point in moduli space. It is reasonable to hope that these integral equations may 
clarify other aspects, both physical and mathematical, of the moduli spaces $\CM$. 

In this note we will take a careful limit of the integral equations, akin to the conformal limit in the 
Thermodynamic Bethe Ansatz literature, and interpret the result as a detailed description of a 
specific  complex Lagrangian sub-manifold $\CL_\epsilon$ in $\CM$. We will argue that the manifold 
coincides with a manifold defined in \cite{Nekrasov:2010ka} by the compactification of the four-dimensional theory 
on a $\Omega_\epsilon$-deformed cigar. 

For theories of the class $\CS$, the relevant sub-manifold $\CL_\epsilon$ is conjecturally the oper manifold. 
The ambient space $\CM$ can be interpreted as a space of flat connections and the oper manifold consists of 
connections which can be gauged into the form of a single a Schr\"odinger-like differential operator on the Riemann surface, 
or a higher-rank generalization of that notion. Thus the TBA equations in the conformal limit 
characterizes the space of opers. Our results thus include and extend previous efforts to use TBA-like equations to 
solve the Schr\"odinger equation with simple potentials \cite{Dorey:1998pt,Bazhanov:1998wj,Dorey:2007zx,Dorey:2007wz}. 
Our methods essentially reconstructs the solutions of a Schr\"odinger equation with rational potential 
from their analytic behavior in the $\hbar \equiv \epsilon$ plane. 

The oper manifold also controls the semiclassical behaviour of conformal blocks for Virasoro or W-algebras. 
In particular, our method should allow the calculation of the semiclassical limit of conformal blocks which are not computable at this moment, 
such as the three-point function of non-degenerate vertex operators. It is natural to wonder if 
our TBA-like equations could somehow be ``quantized'', and compute the full conformal blocks. 

The generating function of the $\CL_\epsilon$ manifold in appropriate (Fenchel-Nielsen) coordinates 
should coincide with the effective superpotential of the two-dimensional gauge theory which emerges 
from the $\Omega$-deformation in a single plane of the four-dimensional gauge theory, as in \cite{Nekrasov:2009rc,Nekrasov:2011bc}. 
Although TBA-like integral equations have appeared in that context as well
\cite{Nekrasov:2009rc}, they appear to be unrelated to the ones presented here. 

\subsection{Outline of our method}
The manifold $\CM$ is defined by a supersymmetric compactification of the four-dimensional gauge theory 
on a circle of radius $R$. It is a hyperk\"ahler manifold, with a $CP^1$ worth of complex structures
which we parameterize by a variable $\zeta$. At $\zeta=0$, the manifold is a complex integrable system, 
a torus fibration over a middle-dimensional base $\CB$, which coincides with the Coulomb branch of the four-dimensional theory. 
The torus fibre is parameterized by the choice of electric and magnetic Wilson lines on the circle. 
The metric on $\CM$ depends on $R$, on the four-dimensional gauge couplings and complex mass parameters $m$ and on the Wilson lines $m_3$ 
for the corresponding flavour symmetries. From this point on, the flavor Wilson lines will be turned off. 

At $\zeta=0$, the manifold is endowed with a canonical complex Lagrangian submanifold $\CL$, a section of the torus fibration, which 
is intuitively defined as the locus where the gauge Wilson lines are turned off. Physically, this submanifold can be defined more precisely by 
the twisted compactification of the four-dimensional theory on a cigar geometry, 
to define a boundary condition $\CL$ for the three-dimensional sigma model on $\CM$ \cite{Nekrasov:2010ka} \footnote{The twist uses $SU(2)_R$. 
As a consequence, the setup is consistent if we define $\CM$ by the supersymmetric compactification on a circle 
which uses a $SU(2)_R$ generator $I_3$ to define the fermion number $(-1)^{I_3}$. Thus $\CM$ is the manifold denoted as $\tilde \CM$ in \cite{Gaiotto:2010be}}.
With the help of the TBA-like integral equations from \cite{Gaiotto:2008cd}, we can describe the manifold $\CL$ 
using complex coordinates for a generic complex structure $\zeta$. Of course, in a generic complex structure for $\CM$, $\CL$ is neither 
a complex, nor a Lagrangian submanifold. 

The TBA equations have an interesting scaling limit, 
the ``conformal limit'', where one sends $R$ and $\zeta$ to zero, keeping $\epsilon = \zeta/R$ fixed. 
If we only focus on the description of $\CM$ as a complex symplectic manifold in complex structure $\zeta$, 
the conformal limit is well defined. Indeed, the complex symplectic structure only depends on 
the radius and $\zeta$ through the mass parameters
\begin{equation}
\mu = \exp \frac{R m}{\zeta} + R \bar m \zeta
\end{equation}
which have a good scaling limit. 
\begin{equation}
\mu = \exp \frac{m }{\epsilon}
\end{equation}
We denote the resulting complex symplectic manifold as $\CM_\epsilon$. 

We will show that something surprising happens to the image $\CL_\epsilon$ of $\CL$ in $\CM_\epsilon$: the scaling limit makes it into a complex Lagrangian submanifold. 
We conjecture that the complex Lagrangian submanifold  $\CL_\epsilon$ is associated to the boundary condition defined by the compactification of the 
four-dimensional theory on an $\Omega_\epsilon$-deformed cigar. We do not offer a full proof of this conjecture, though it can be motivated 
in part by the analysis of \cite{Nekrasov:2010ka}. Instead, we will attempt to demonstrate directly that for theories in the class $\CS$, the 
manifold $\CL_\epsilon$ is the oper manifold. 

This is possible thanks to a second set of integral equations \cite{Gaiotto:2011tf} which compute directly the solutions
of Hitchin's equations. We can show explicitly how the solutions associated to points in $\CL$ 
go to opers in the conformal limit. 

The main assumption in this paper is that the solutions of the integral equations have a good conformal limit, which is reasonably well-behaved 
at large $\epsilon$. As systematic tests of this assumption would require extensive numerical work, we leave them to future publications. 
We will limit ourselves here to simple examples. 

\section{General considerations}
Our starting point are the TBA-like equations used to describe the 
metric on the moduli space $\CM$ of four-dimensional ${\cal N}=2$ gauge theories compactified on a circle. 
\begin{equation} \label{eq:int-old}
\log X_{\gamma}(\zeta) = \frac{Z_\gamma}{\zeta} + \I \theta_\gamma + \bar Z_\gamma \zeta +  \sum_{\gamma'} \omega(\gamma', \gamma) \frac{1}{4 \pi \I} \int_{\ell_{\gamma'}} \frac{\de \zeta'}{\zeta'} \frac{\zeta' + \zeta}{\zeta' - \zeta} \log(1 - \sigma(\gamma')X_{\gamma'}(\zeta')).
\end{equation}
with reality condition $X_{\gamma}(\zeta)=\overline{X_{-\gamma}\left(-1/\bar \zeta\right)}$. 

We lightened a bit the notation compared to the reference \cite{Gaiotto:2008cd}. For the purpose of this note, 
we will not need to review the somewhat intricate geometric meaning of the symbols $\gamma$, $Z_\gamma$, $\theta_\gamma$, $ \omega(\gamma', \gamma)$, etc. 
Roughly, the charge $\gamma$ labels a certain choice of complex coordinates $X_\gamma$. 
The periods $Z_\gamma$ label a point on the basis of the complex integrable system, 
and the angles $\theta_\gamma$ the fibre. 
The canonical integration contours 
$\ell_\gamma$ in the $\zeta$ plane are such that $\frac{Z_\gamma}{\zeta}$ is real and negative. We will discuss alternative choice of contours at the end of this section.

The hyperk\"ahler metric on $\CM$ is computed by plugging the solutions into a complex symplectic form $\Omega(\zeta) = \langle d \log X, d \log X \rangle$.
The asymptotic behaviour at small and large $\zeta$, together with the identity $\langle d Z, d Z \rangle=0$, imply
\begin{equation}
\Omega(\zeta) = \frac{\omega^+}{\zeta} + \omega^3 + \omega^- \zeta.
\end{equation}
and thus we arrive to the complex symplectic form $\omega^+$ and K\"ahler form $\omega^3$ in complex structure $\zeta=0$. 
The BPS degeneracies $\omega(\gamma', \gamma)$ are determined by the requirement that the metric should be smooth.

We are interested in a special section $\CL$ of the moduli space, which is defined by setting the angles $\theta_\gamma$ to zero. 
This statement requires a little clarification, as the formalism allows for sign redefinitions of the $X_\gamma$ functions, 
which lead to shifts of the $\theta_\gamma$ by multiples of $\pi$, and changes in the choice of the ``quadratic refinement''
$\sigma(\gamma)$. There is a canonical choice provided by the gauge theory \cite{Gaiotto:2008cd,Gaiotto:2011tf} which is also mathematically natural. \footnote{
The canonical quadratic refinement is the difference between the $SU(2)_R$ ``fermion number'' $(-1)^{I_3}$ of a BPS particle of charge $\gamma$ and the more standard fermion number $(-1)^{J_3}$ defined through the angular momentum generator. }

Once we restrict $\theta_\gamma$ to zero, the equations and their solutions gain an extra $Z_2$ symmetry
$X_{-\gamma}(- \zeta) = X_\gamma(\zeta)$ which allows one to combine together the contributions from $\pm \gamma'$ in the sum. 
\begin{equation} \label{eq:int-old-half}
\log X_{\gamma}(\zeta) = \frac{Z_\gamma}{\zeta} + \bar Z_\gamma \zeta +  \sum_{\gamma'>0} \omega(\gamma', \gamma) \frac{\zeta}{\pi \I} \int_{\ell_{\gamma'}} \frac{\de \zeta'}{(\zeta')^2 - (\zeta)^2} \log(1 - \sigma(\gamma')X_{\gamma'}(\zeta')).
\end{equation}
The cancellation of the order $1$ terms in the $\zeta \to 0$ expansion has an important consequence: 
if we plug the expansion in the expression for the complex symplectic form 
\begin{equation}
\Omega(\zeta) = \langle d \log X, d \log X \rangle
\end{equation}
we verify that we are describing a complex Lagrangian section $\CL$ in complex structure $\zeta=0$. 

Next, we would like to take the ``conformal limit`` in the TBA: introduce a radius parameter by the rescaling $Z_\gamma \to R Z_\gamma$, 
and send $R$ to zero. Because the new integration kernel goes to zero if $\zeta/\zeta'$ is very different from $1$, 
we can self-consistently focus on the behaviour of the functions at small $\zeta$, by setting $\zeta = R \epsilon$,$\zeta' = R \epsilon'$ and keeping $\epsilon$ 
fixed as $R$ is sent to zero. 

The resulting set of equations take the simpler form:
\begin{equation} \label{eq:int-new}
\log X_{\gamma}(\epsilon) = \frac{Z_\gamma}{\epsilon} +  \sum_{\gamma'>0} \omega(\gamma', \gamma) \frac{\epsilon}{\pi \I} \int_{\ell_{\gamma'}} \frac{\de \epsilon'}{(\epsilon')^2 - (\epsilon)^2} \log(1 - \sigma(\gamma')X_{\gamma'}(\epsilon')).
\end{equation}
The functions $X_{\gamma}(\epsilon)$ live in a complex manifold $\CM_\epsilon$ defined by the limiting value of the mass parameters 
\begin{equation}
\log \mu = \frac{m}{\epsilon}
\end{equation}

\subsection{Large $\epsilon$ behaviour}
Clearly, we are making the assumption that the solutions of the integral equations have a good conformal limit, and remain somewhat well- 
behaved at large $\epsilon$. This assumption is crucial for our results to hold. We can gain some insight if we recall the detailed analysis of 
\cite{Gaiotto:2008cd}. 

For that purpose, it is useful to reintroduce the angles $\I \theta_\gamma$ in the conformal limit. More precisely, one can massage the integration kernels a little bit and introduce a complexified version $\theta^+_\gamma$ of the angles
\begin{equation} \label{eq:int-old2}
\log X_{\gamma}(\zeta) = \frac{R Z_\gamma}{\zeta} + \I \theta^+_\gamma + R \bar Z_\gamma \zeta + \zeta \sum_{\gamma'} \omega(\gamma', \gamma) \frac{1}{2 \pi \I} \int_{\ell_{\gamma'}} \frac{\de \zeta'}{\zeta'} \frac{1}{\zeta' - \zeta} \log(1 - \sigma(\gamma')X_{\gamma'}(\zeta')).
\end{equation}
Then we can take the conformal limit
\begin{equation} \label{eq:int-new2}
\log X_{\gamma}(\epsilon) = \frac{Z_\gamma}{\epsilon} + \I \theta^+_\gamma +\epsilon \sum_{\gamma'} \omega(\gamma', \gamma) \frac{1}{2 \pi \I} \int_{\ell_{\gamma'}} \frac{\de \epsilon'}{\epsilon'} \frac{1}{\epsilon' - \epsilon} \log(1 - \sigma(\gamma')X_{\gamma'}(\epsilon')).
\end{equation}

If we keep the angles, we can look at certain differential equations in $R$ and $\zeta$ satisfied by the solutions $X_{\gamma}(\zeta)$,
which have an irregular singularity at both $\zeta=0$ and $\zeta = \infty$, see section $5.5$ of \cite{Gaiotto:2008cd}.
These equations can be combined into a differential equation in $\epsilon$, which has an irregular singularity 
as $\epsilon \to 0$, but only a regular singularity as $\epsilon \to \infty$
\begin{equation}
\epsilon\, \partial_\epsilon  X_{\gamma} = \left[- \I \frac{Z }{\epsilon} + c\right] \cdot \partial_{\theta^+}X_{\gamma} 
\end{equation} 
for some $\epsilon$-independent functions $c_\gamma$.

The regular singularity suggests that the $X_{\gamma}(\epsilon)$ will have a power-law behaviour at large $\epsilon$. 
The monodromy at large $\epsilon$ should coincide with the monodromy around the origin, 
which is decomposed into a product of Stokes factors for the irregular singularity at $\epsilon \to 0$. 
The Stokes factors coincide with the discontinuities of the solutions across the $\ell_\gamma$ rays, 
the KS transformations 
\begin{equation}
X_\gamma \to X_\gamma(1-\sigma(\gamma) X_{\gamma'})^{\omega(\gamma', \gamma)}
\end{equation}
Thus the large $\epsilon$ behaviour is constrained by the BPS spectrum of the theory. 

As an example, suppose that the solutions $X_{\gamma}(\epsilon)$ to go to a constant $X^\infty_\gamma$
as $\epsilon \to \infty$. We can evaluate the integral in the large $\epsilon$ limit by looking at large $\epsilon'$, 
and pulling the $\log$ out of the integral. We get
\begin{equation} 
\log X^\infty_{\gamma} = \sum_{\gamma'>0}\omega(\gamma', \gamma) \frac{1}{2} \sigma_{\gamma, \gamma'}  \log(1 - \sigma(\gamma) X^\infty_{\gamma'}).
\end{equation}
where the sign $\sigma_{\gamma,\gamma'}$ is $+$ if the ray $\ell_\gamma$ lies counterclockwise from the ray $\ell_{\gamma'}$.
We can rewrite this as the algebraic equations
\begin{equation} 
(X^\infty)^2_{\gamma} = \prod_{\gamma'>0} (1 - \sigma(\gamma) X^\infty_{\gamma'})^{\omega(\gamma', \gamma)  \sigma_{\gamma, \gamma'}}.
\end{equation}

Depending on the model, these equations may have isolated solutions, or a moduli space of solutions. In the latter case, we will pick a specific choice as we vary 
the $Z_\gamma$. Thus, at least locally, the limiting values $X^\infty$ do not depend on the $Z_\gamma$. This has an important consequence: 
if we plug the $X_\gamma(\epsilon)$ in the complex symplectic form $\Omega(\epsilon)$ of $\CM_\epsilon$, 
the $\epsilon \to 0$ expansion and the limiting behaviour at $\epsilon \to \infty$ force $\Omega$ to vanish. 
In other words, we are describing a complex Lagrangian submanifold $\CL_\epsilon$ in $\CM_\epsilon$. 
If the $X_{\gamma}(\epsilon)$ do not go to constants, but rather grow polynomially in $\epsilon$ at large $\epsilon$, we can reach a similar conclusion, as long as 
the leading growth can be chosen to be $Z_\gamma$-independent.

\subsection{Spectrum generator}
Notice that the integration contours $\ell_\gamma$ can be deformed rather freely inside a half-plane centred on their original position, 
as long as their relative order is preserved. If several rays are collapsed together, we can still use similar integral equations, but we need to 
combine the discontinuities across the rays properly. In the most extreme case, all the rays inside a half-plane can be collapsed together,
leaving a single integration contour. 

If we define the ``spectrum generator'' as the composition of all the discontinuities \cite{Gaiotto:2009hg,Caetano:2012ac}, written as the coordinate transformation relating the 
$X_\gamma$ on the two sides of the cut, 
\begin{equation}
X^+_\gamma = X^-_\gamma F(X^-)
\end{equation}
then the integral equations take the form (beyond the conformal limit, this form of the equations was used by \cite{Caetano:2012ac}) 
\begin{equation} \label{eq:int-old-F}
\log X_{\gamma}(\epsilon) = \frac{Z_\gamma}{\epsilon} + \frac{\epsilon}{\pi \I} \int_{\ell_{\gamma'}} \frac{\de \epsilon'}{(\epsilon'- i 0)^2 - (\epsilon)^2} \log F(X(\epsilon')).
\end{equation}
This can be very useful, as the spectrum generator for a theory is much simpler to obtain than the individual BPS degeneracies $\omega(\gamma', \gamma)$. 
The $i0$ prescription is needed because the discontinuity of a coordinate depends on the coordinate itself once rays are collapsed together. 

In this form, the behaviour at large $\epsilon$ is simpler to understand. The equations reduce to 
\begin{equation}
(X^\infty)^2_{\gamma} F(X^\infty)=1
\end{equation}
i.e. the spectrum generator must send $X^\infty_\gamma$ to $X^\infty_{-\gamma}$.

\section{Simple examples}
All of our basic examples will be taken from $\CS[A_1]$ theories, so that the moduli space $\CM_\epsilon$ is a space of monodromy data for complex $SL(2)$ flat connections. 
We want to verify that $\CL_\epsilon$ coincides with the monodromy data of $SL(2)$ opers. More precisely, we anticipate opers of the form
\begin{equation}
-\partial_z^2 + \frac{\phi(z)}{\epsilon^2} + t_0(z)
\end{equation}
where $\phi(z)$ is a quadratic differential such that the $Z_\gamma$ are periods of $\sqrt{\phi(z)}$ and $t_0(z)$ is a classical stress tensor 
determined somehow by the large $\epsilon$ behaviour of the TBA equations. 

As for the original $SL(2)$ Hitchin system, the $X_\gamma(\epsilon)$ variables coincide with ``Fock coordinates'', cross-ratios of Wronskians 
of certain ``small solutions'' $s_a$, which are solutions of the Schr\"odinger differential equation with prescribed behaviour at singularities. 
We are only allowed to use Wronskians which can be estimated in the $\epsilon \to 0$ limit by a WKB approximation, 
which turn out to correspond to the edges of a ``WKB triangulation''. Each triangle is centred around a turning point $\phi(z)=0$, 
and each edge $E$ is associated to a compact cycle $\gamma_E$ on the spectral curve
\begin{equation}
x^2 = \phi(z),
\end{equation} which is the charge which labels 
the corresponding cross-ratio $X_E \equiv X_{\gamma_E}$. Indeed, by the WKB approximation, 
the asymptotic behaviour of $X_E$ is $Z_{\gamma_E}$, the period of $\lambda = x dz$ on $\gamma_E$. 

We will also find it useful to look directly at the Wronskians themselves, $T_E \equiv X_{\hat \gamma_E}$, whose asymptotics 
are controlled by the periods $Z_{\hat \gamma_E}$ of $\lambda$ on non-compact cycles $\hat \gamma_E$, which coincide with the edges themselves. 
The non-compact cycles $\hat \gamma_E$ form a dual basis to the cycles $\gamma_E$, and we expect the Wronskians 
to be computed by the same integral equation, with an appropriate choice of $\omega(\gamma, \hat \gamma')$. 
The spectrum generator transformation is easily extended to the Wronskians. 

In this section we will focus at first on a handful of ``local'' examples, where the Schr\"odinger equation can be exactly solved. 
The analytic calculations will use the following two definite integrals: for positive real part of $x$, 
\begin{equation}
\log \Gamma(\frac{1}{2} + x) = x (-1+\log x )+ \log \sqrt{2 \pi}+\frac{1}{\pi} \int_0^\infty \frac{d t}{t^2+1}\log \left(1+e^{- 2 \pi \frac{x}{t}} \right)
\end{equation}
and
\begin{equation}
\log \Gamma(x) = x (-1+\log x )+ \log \sqrt{\frac{2 \pi}{x}}+ \frac{1}{\pi} \int_0^\infty \frac{d t}{t^2+1}\log \left(1-e^{- 2 \pi \frac{x}{t}} \right)
\end{equation}

Later in the section, we will make numerical comparisons for more general choices of spectral curve and Schr\"odinger operator. 
For numerical calculations, and for comparison to the integral equations in this and the next section, it is very useful to use a modified 
form of the Schr\"odinger equation. Starting from 
\begin{equation}
\left[ -\partial_z^2 + \frac{\phi(z)}{\epsilon^2} + t_0(z) \right] \psi(z)=0
\end{equation}
and writing the wave-function $\psi(z) = e^{\frac{1}{\epsilon}\int^z x dz} f(z)$ we get 
\begin{equation} \label{eq:mod}
\left[ -\partial_z^2 - \frac{2 x}{\epsilon} \partial_z - \frac{\partial_z x}{\epsilon} + t_0(z) \right] f(z)=0
\end{equation}

This differential equation can be easily integrated numerically along the paths $\hat \gamma$. The combination $\sqrt{x} f(z)$ 
has a finite limit as we go to infinity, and the ratio of $\sqrt{x} f(z)$ at the end-points of the path can be readily compared with 
the non-trivial part of the Wronskians, $e^{- \frac{Z_{\hat \gamma}}{\epsilon}} X_{\hat \gamma}$. 

\subsection{The harmonic oscillator}
There is a simple setup, which works as a local model for the metric on $\CM$ near a singular locus of the Coulomb branch, 
where a single BPS hypermultiplet becomes massless. It is associated to the $A_1$ spectral curve with a rank $2$ irregular singularity 
at infinity, denoted as $AD_2$ in \cite{Gaiotto:2009hg}.
\begin{equation}
x^2 = z^2+ 2 a
\end{equation}

Physically, this is associated to the theory of a single BPS hypermultiplet, of mass $Z_e = 2 \pi i a$.
The period $Z_e$ is the period of the differential $\lambda = x dz$ along the finite cycle $\gamma_e$ surrounding the origin at large $z$.
The corresponding coordinate is uncorrected, $X_e = \exp \frac{2 \pi \I a}{\epsilon}$. 

We can define a dual, non-compact cycle $\gamma_m = \hat \gamma_e$ lying on the real axis. To make this statement and subsequent formulae precise, it is useful to keep 
$a$ close to the positive real axis. Analytic continuation to other values of $a$ is straightforward. The corresponding (regularized) period is
\begin{equation}
Z_m = \Lambda^2 + a \left(1 - \ln \frac{a}{2\Lambda^2} \right)
\end{equation}
This controls the asymptotics of the T-function $T_e$ dual to $X_e$, which we will denote as $X_m$. 
We can compute $X_m$ right away from the integral equation, using $\omega(e, m) = 1$ and $\sigma(e)=-1$,
\begin{equation} \label{eq:Xm}
\log X_m = \frac{Z_m}{\epsilon} + \frac{\epsilon}{\pi \I} \int_{\ell_{-\gamma_e}} \frac{\de \epsilon'}{(\epsilon')^2 - (\epsilon)^2} \log(1 +e^{-\frac{2 \pi \I a}{\epsilon'}})
\end{equation}
to obtain the analytic form
\begin{align}
X_m =  X_m^+ &= e^{\Lambda^2/\epsilon}\left(2\Lambda^2/\epsilon \right)^{a/\epsilon} \frac{\sqrt{2\pi } }{\Gamma\left(\frac{1}{2}+\frac{a}{\epsilon}\right)}\qquad &\mathrm{Re} a/\epsilon>0 \cr 
X_m = X_m^- &=e^{\Lambda^2/\epsilon}\left(-2 \Lambda^2/\epsilon \right)^{a/\epsilon}  \frac{ \Gamma\left(\frac{1}{2}-a\right)}{\sqrt{2 \pi }} \qquad &\mathrm{Re} a/\epsilon<0
\end{align}
A basic check is to verify the discontinuity $X_m^+ = X_m^- (1 +e^{\pm \frac{2 \pi \I a}{\epsilon'}})$ along the positive or negative imaginary $a/\epsilon$ axis. 

We can compare these functions with the analogous quantities computed from the harmonic oscillator Schr\"odinger operator 
\begin{equation}
- \epsilon^2 \partial_z^2 \psi+ (z^2 + 2 a) \psi=0
\end{equation}
If we look at a solution which decreases along the positive real axis as 
\begin{equation}
\psi_R \sim e^{\frac{\Lambda^2 - z^2}{2 \epsilon}}
   \left(L/z\right)^{a/\epsilon}\sqrt{\frac{\epsilon}{2 z}}
\end{equation}
and take the Wronskian with a solution which decreases along the negative real axis as 
\begin{equation}
\psi_L \sim e^{\frac{\Lambda^2 - z^2}{2 \epsilon}}
   \left(-L/z\right)^{a/\epsilon}\sqrt{-\frac{\epsilon}{2 x}}
\end{equation}
we obtain $X^+_m$.
Similarly $(X_m^-)^{-1}$ arises from the Wronskian of wavefunctions which decrease along the imaginary axis $\psi_U$ and $\psi_D$.
 This agrees with the definition given for the $AD_2$ theory in \cite{Gaiotto:2009hg}.
 
The comparison can be made less tedious by using \ref{eq:mod}, and comparing directly the result of the integral in \ref{eq:Xm} 
\begin{equation} \label{eq:Xm-again}
\frac{\epsilon}{\pi \I} \int_{\ell_{-\gamma_e}} \frac{\de \epsilon'}{(\epsilon')^2 - (\epsilon)^2} \log(1 +e^{-\frac{2 \pi \I a}{\epsilon'}})
\end{equation}
with the change in $\log \sqrt{x} f(z)$ along the open path $\gamma_m$. 

\subsection{A more intricate local model}
Our next model controls the behavior of moduli spaces of $\CS[A_1]$ theories 
as the mass parameter for an $SU(2)$ flavor symmetry is turned off,
near the locus in the Coulomb branch where a Higgs branch opens up. 

The spectral curve for the model is
\begin{equation}
x^2 = 1 + \frac{2 a}{z} + \frac{c^2}{z^2}
\end{equation}
This curve has two finite cycles, corresponding to 
\begin{equation}
Z_1 = 2 \pi i (c+a) \qquad Z_2 = 2 \pi i (c-a).
\end{equation}
We define $Z_{1+2}= 4 \pi i c$.

The corresponding basis of non-compact cycles runs on the positive real axis and on the negative real axis respectively
(say with $a$ and $c$ real and positive), 
giving 
\begin{align}
Z_{\hat 1} = \Lambda + (c+a) \left(1 - \ln \frac{c+a}{2 \Lambda} \right)-2 c \left(1 - \ln \sqrt{\frac{\tilde \Lambda}{\Lambda}} c\right) \cr
Z_{\hat 2} = \Lambda + (c-a) \left(1 - \ln \frac{c-a}{2 \Lambda} \right)-2 c \left(1 - \ln \sqrt{\frac{\tilde \Lambda}{\Lambda}} c\right)
\end{align}

The coefficients of the logarithmic singularities matches the non-zero 
\begin{equation} \omega(1, \hat 1) = - \omega(1+2, \hat 1)=1 \qquad \qquad  \omega(2, \hat 2) = - \omega(1+2, \hat 2)=1,
\end{equation} and the corresponding T-functions can be computed right away in analytic form (we also need $\sigma(1) = \sigma(2) = -\sigma(1+2) =-1$). 
For example,
\begin{align}
X_{\hat 1} = e^{\Lambda/\epsilon}\left(2\Lambda/\epsilon \right)^{a/\epsilon}\left(\tilde \Lambda\right)^{c/\epsilon} \frac{\sqrt{2c/\epsilon} \Gamma\left(\frac{2c}{\epsilon}\right)}{\Gamma\left(\frac{1}{2}+\frac{c+a}{\epsilon}\right)}\qquad &\mathrm{Re} (c+a)/\epsilon>0 \, \mathrm{Re} c/\epsilon>0\end{align}
This coincides with an appropriate Wronskian 
of sections for the oper
\begin{equation}
-\epsilon^2 \partial_z^2 + 1 + \frac{2 a}{z} + \frac{c^2-\frac{\epsilon^2}{4}}{z^2}
\end{equation}
with behaviour
\begin{equation}
 \sqrt{\epsilon z/(2c)} \left(\frac{2z}{\epsilon \tilde \Lambda}\right)^{c/\epsilon}
\end{equation}
near the origin and 
\begin{equation}
\sqrt{\epsilon/2} e^{-\frac{x}{\epsilon}} (x/\Lambda)^{-a/\epsilon}
\end{equation}
at positive infinity.

It is interesting to observe that the $-\frac{\epsilon^2}{4}$ correction at the regular singularity is just what one would expect from the semi-classical limit 
of the AGT dictionary \cite{Alday:2009aq}. There is a simple interpretation 
for the singular behavior of the sub-leading stress-tensor
\begin{equation}
t_0(z) = -\frac{1}{4 z^2}
\end{equation}
If we do a conformal transformation $z = e^s$ to make the regular puncture into a tube, 
$t_0$ disappears. This pattern will persist in other examples.  

A simplified version of this local system with a rank $1/2$ irregular singularity at infinity 
\begin{equation}
x^2 = \frac{1}{z} + \frac{c^2}{z^2}
\end{equation}
with a single $\omega(e, m)=-1$, $\sigma(e)=1$ state can be similarly matched to the oper
\begin{equation}
-\epsilon^2 \partial_z^2 + \frac{1}{z} + \frac{c^2-\frac{\epsilon^2}{4}}{z^2}
\end{equation}

In both problems, we can dispense of the need of carefully regulating the open periods if we base our comparisons on \ref{eq:mod}.
\subsection{Three-punctured sphere}
The $A_1$ example with three regular punctures is more elaborate, but predictable.
The spectral curve is 
\begin{equation}
x^2 = \frac{-{c_0}^2-{c_1}^2+{c_\infty}^2}{(z-1)
   z}+\frac{{c_0}^2}{z^2}+\frac{{c_1}^2}{(z-1)^2}
\end{equation}
There are three natural cycles of periods
\begin{align}
Z_1 &= 2 \pi i (-c_0 + c_1 + c_\infty) \cr
Z_2 &= 2 \pi i (c_0 - c_1 + c_\infty) \cr
Z_3 &= 2 \pi i (c_0 + c_1 - c_\infty) 
\end{align}
and BPS degeneracies 
\begin{align}
\omega(1, \hat 1) = - \omega(1+2, \hat 1)=-\omega(1+3, \hat 1) = \omega(1+2+3, \hat 1) =1 \cr
\omega(2, \hat 2) = - \omega(1+2, \hat 2)=-\omega(2+3, \hat 2) = \omega(1+2+3, \hat 2) =1 \cr
\omega(3, \hat 3) = - \omega(2+3, \hat 3)=-\omega(1+3, \hat 3) = \omega(1+2+3, \hat 3) =1 
\end{align}
and 
\begin{equation}
\sigma(1) = \sigma(2) = \sigma(3) = - \sigma(1+2) = - \sigma(1+3) = - \sigma(2 + 3) = \sigma(1+2+3) =-1.
\end{equation}

A tedious but straightforward calculation produces T-functions which precisely match the Wronskians
of the corresponding sections of 
\begin{equation}
-\epsilon^2 \partial_z^2 + \frac{-{c_0}^2-{c_1}^2+{c_\infty}^2+\epsilon^2/4}{(z-1)
   z}+\frac{{c_0}^2-\epsilon^2/4}{z^2}+\frac{{c_1}^2-\epsilon^2/4}{(z-1)^2}
\end{equation}
\subsection{The cubic}
The first example where numerical calculations are needed is the $AD_3$ theory,  
\begin{equation}
x^2 = z^3 + \Lambda z + u
\end{equation}

In order to carry on the calculation efficiently, we can first specialize to a particularly symmetric point, $u=0$. With no loss of generality, we can set the scale $\Lambda$ to $1$. 
\begin{equation}
x^2 = z^3 + z
\end{equation}
The two basic periods 
\begin{equation}
Z_1 = \frac{8 \sqrt{2} \pi ^{3/2}}{5 \Gamma
   \left(\frac{1}{4}\right)^2}e^{\I \pi/4} \qquad Z_2 = \frac{8 \sqrt{2} \pi ^{3/2}}{5 \Gamma
   \left(\frac{1}{4}\right)^2}e^{3 \I \pi/4} 
\end{equation}
satisfy a relation $Z_2 = \I Z_1$, which implies an enhancement of the usual $Z_2$ symmetry to $Z_4$: 
$X_2(\I \epsilon) = X_1(\epsilon)$. The non-zero degeneracies are   $\omega(1,2) =-\omega(2,1) = 1$ with $\sigma(1)=\sigma(2)=-1$
and thus we can collapse the integral equations to 
\begin{equation} \label{eq:int-cubic}
\log X_1(\epsilon) = \frac{Z_1}{\epsilon} + \frac{\epsilon}{\pi} \int_{\ell_1} \frac{\de \epsilon'}{(\epsilon')^2 +(\epsilon)^2} \log(1 +X_1(\epsilon')).
\end{equation}
We can readily solve the equation numerically, by iterating it a few times. 
The solution goes to the golden ratio at infinity, as we have
\begin{equation}
(X^\infty_1)^2 = 1+X^\infty_1
\end{equation}

The functions $\log X_{1,2}(\epsilon)$, computed numerically on the real positive $\epsilon$ axis, 
agree with high numerical precision with the appropriate transport coefficients of the 
Schr\"odinger operator 
\begin{equation}
- \epsilon^2 \partial_z^2 + z^3 + z
\end{equation}

\subsection{Pure $SU(2)$}
The next example is associated to the spectral curve
\begin{equation}
x^2 = \frac{1}{z} + \frac{u}{z^2} + \frac{1}{z^3}
\end{equation}
If we set $u=0$, we gain the same sort of $Z_4$ symmetry as for the cubic example. The reduced integral equation is almost 
identical: it differs by a crucial factor of $2$: 
\begin{equation} \label{eq:int-su}
\log X_1(\epsilon) = \frac{Z_1}{\epsilon} + 2 \frac{\epsilon}{\pi} \int_{\ell_1} \frac{\de \epsilon'}{(\epsilon')^2 +(\epsilon)^2} \log(1 +X_1(\epsilon')).
\end{equation}

This factor of $2$ has deep consequences. The solutions cannot asymptote to a constant at infinity: 
\begin{equation}
X^\infty_1 = 1+X^\infty_1
\end{equation}
does not make sense. 
Indeed, it is easy to argue that a logarithmic divergence $k \log \epsilon$ of $\log X_1$ is self-consistent for 
any $k$. The integral equations in the conformal limit do not fix $k$. 

On the other hand, if we look at the full integral equations, and take the conformal limit of their solutions, 
the limit appears well-defined. The limiting solutions appear to diverge roughly as $\frac{1}{2} \log \epsilon$, 
i.e. $X_1$ appears to grow as $\sqrt{\epsilon}$. 

Some educated guesswork produces a candidate Schr\"odinger operator whose transport coefficients 
appear to match numerically the conformal limit of the solutions of the full integral equations at $u=0$:
\begin{equation}
- \epsilon^2 \partial_z^2 +  \frac{1}{z} + \frac{u-\epsilon^2/4}{z^2} + \frac{1}{z^3}
\end{equation}

\section{Specialization to Hitchin moduli space}
If we are dealing with a theory in class $\CS$, so that $\CM$ is roughly the Hitchin's moduli space, we can use a further set of integral equations which 
produce directly the flat sections of the auxiliary flat connection 
\begin{equation}
\CA = \frac{\Phi}{\zeta} + A + \zeta \bar \Phi
\end{equation}
for the Hitchin system on a punctured Riemann surface $C$, and thus the solution $A,\Phi$ of the Hitchin system labeled by a point in $\CM$.
Then we can verify directly if the equations provide flat sections of opers in the conformal limit.  

We first introduce the auxiliary functions $x_{\gamma_{ij'}}$, labeled by open paths $\gamma_{ij'}$ on the spectral curve
\begin{equation} \label{eq:spec}
\det  \left[ x dz - \Phi(z)\right] =0,
\end{equation} through the equation 
\begin{equation} \label{eq:int-old-x}
\log x_{\gamma_{ij'}}(\zeta) = \frac{Z_{\gamma_{ij'}}}{\zeta} + \I \theta_{\gamma_{ij'}} + \bar Z_{\gamma_{ij'}} \zeta +  \sum_{\gamma'} \omega(\gamma', \gamma_{ij'}) \frac{1}{4 \pi \I} \int_{\ell_{\gamma'}} \frac{\de \zeta'}{\zeta'} \frac{\zeta' + \zeta}{\zeta' - \zeta} \log(1 - X_{\gamma'}(\zeta')).
\end{equation}
The coefficients $\omega(\gamma', \gamma_{ij'})$ are piecewise constant on $C$, thus the dependence 
on the initial and final points of $\gamma_{ij'}$ is locally captured by the periods $Z_{\gamma_{ij'}}$ of the canonical differential 
$\lambda = x dz$ on the spectral curve. The labels $i$ and $j'$ indicate on which sheets of the spectral curve the path $\gamma_{ij'}$ ends. 

Then we define a matrix 
\begin{equation} \label{eq:int-2}
g_k(\zeta) = g_k^0 + \sum_{\ell \neq k, \gamma_{\ell k}} \mu(\gamma_{\ell k})
\frac{1}{4 \pi \I} \int_{\ell_{\gamma_{\ell k }}} \frac{\de \zeta'}{\zeta'}
\frac{\zeta' + \zeta}{\zeta' - \zeta} g_\ell (\zeta') x_{\gamma_{\ell k}}(\zeta'),
\end{equation}
and its inverse 
\begin{equation} \label{eq:int-2-inv}
g_{-k}(\zeta) = g_{-k}^0 - \sum_{\ell \neq k, \gamma_{  k \ell}} \mu(\gamma_{  k \ell} )
\frac{1}{4 \pi \I} \int_{\ell_{\gamma_{  k \ell} }} \frac{\de \zeta'}{\zeta'}
\frac{\zeta' + \zeta}{\zeta' - \zeta} x_{\gamma_{  k \ell}}(\zeta') g_{-\ell} (\zeta')  .
\end{equation}

These matrices define a local gauge transformation which diagonalizes the Hitchin complex flat connection $\CA$
and reduces it to an abelian connection whose local holonomies are encoded in the $x_{\gamma_{ij'}}$. 
The source terms $g_k^0$ and $g_{-k}^0$ pick a specific choice of gauge for the solution of the Hitchin system. We will come
back to them momentarily. 

Notice that the abelian connection can be recovered simply as $a_{j'} = d x_{\gamma_{ij'}}$,
where $d$ acts on the endpoint of the path $\gamma_{ij'}$.
We can write directly 
\begin{equation} \label{eq:dint-old}
a_i(\zeta) = \frac{\lambda_i}{\zeta} + \I d\theta_i + \bar \lambda_i \zeta +  \sum_{\gamma'} d\omega(\gamma', i)  \frac{1}{4 \pi \I} \int_{\ell_{\gamma'}} \frac{\de \zeta'}{\zeta'} \frac{\zeta' + \zeta}{\zeta' - \zeta} \log(1 - X_{\gamma'}(\zeta')).
\end{equation}
The closed forms $d\omega(\gamma', i)$ will be supported on specific codimension $1$ loci on $C$. 

The complex flat connection is recovered as 
\begin{equation}
\CA = \sum_i  g_i a_i g_{-i} + g_i dg_{-i} 
\end{equation}
It is useful to expand everything around $\zeta=0$. We can pick a convenient complexified gauge choice 
by rewriting the $g_k$ integral equation as 
\begin{equation} \label{eq:int-2-gauge}
g_k(\zeta) = g_k^+ + \sum_{\ell \neq k, \gamma_{\ell k}} \mu(\gamma_{\ell k})
\frac{\zeta}{2 \pi \I} \int_{\ell_{\gamma_{\ell k }}} \frac{\de \zeta'}{\zeta'}
\frac{1}{\zeta' - \zeta} g_\ell (\zeta') x_{\gamma_{\ell k}}(\zeta'),
\end{equation}
and similarly for $g_{-i}$. 

If we expand 
\begin{equation}
g_i = g_i^+ + \cdots \qquad g_{-i} = g^+_{-i} + \cdots
\end{equation}
and
\begin{equation}
a_i = \frac{\lambda_i}{\zeta} + \I d\theta_i + \rho_i +\cdots
\end{equation}
The correction term 
\begin{equation}
\rho_i =  \sum_{\gamma'} d\omega(\gamma', i) \frac{1}{4 \pi \I} \int_{\ell_{\gamma'}} \frac{\de \zeta'}{\zeta'} \log(1 - X_{\gamma'}(\zeta')) 
\end{equation}
is a real form, due to the reality conditions on $X_\gamma$ and $\omega(-\gamma', i) = - \omega(\gamma', i)$. 
Thus we find 
\begin{equation}
\Phi = \sum_i  g^+_i \lambda_i g^+_{-i}  
\end{equation}
and (remember that the canonical one form $\lambda$ is holomorphic)
\begin{equation}
A_{\bar z} = \sum_i  g^+_i (\I d\theta_i + \rho_i )_{(0,1)} g^+_{-i} + g^+_i dg^+_{-i} 
\end{equation}
We can express these relations as 
\begin{equation} \label{eq:eigen}
\Phi g^+_i = \lambda_i g^+_{i}  \qquad \qquad D_{\bar z} g^+_i = (\rho_i +\I d\theta_i)_{(0,1)}  g^+_i
\end{equation}

We would like to match this with the standard parameterization of the Higgs bundle in terms of the spectral curve data.  
The construction is reviewed beautifully in section $4$ of \cite{2007arXiv0710.5939F}, which explains several mathematical subtleties which are 
important in the following analysis. The spectral curve for an $SL(K)$ Hitchin system 
is the curve of eigenvalues of $\Phi$ \ref{eq:spec}.
The Higgs bundle defines a line bundle on the spectral curve, defined as the co-kernel of $x dz - \Phi$. 
The line bundle has a non-trivial Chern class. It can be made into a degree zero line bundle 
by combining it with the difference between the square roots of the canonical bundles of the base curve $C$ 
and the spectral curve $\Sigma$. Essentially, the point is that the eigenline bundle has curvature localized at the 
turning points, where two eigenvalues collide. Then the degree zero line bundle can be made into a 
flat $U(1)$ bundle, and used to parameterize the fibre of Hitchin fibration. 

This is exactly what we see in $\ref{eq:eigen}$! The $g^+_{i}$ intertwine the full bundle $V$ and the eigenline bundles $V_i$.
A local calculation near the turning points show that the $g^+_{i}$ matrix must have a precise singularity there in order to 
have a smooth solution of the integral equations. Roughly, $g_i$ diverges as $\prod_{j \neq i} (x_i - x_j)^{-1/2}$. 
The singularity is exactly such that we can 
reinterpret $g^+_{i}$ as a smooth intertwiner between $V \otimes K_C^{1/2}$ and $V_i \otimes K_\Sigma^{1/2}$.

We see thus that $\partial_{\bar z} -(\rho_i +\I d\theta_i)$ is the Abelian connection on the degree $0$ eigenline bundle. 
We can verify by hand that the forms $\rho_i$ are closed on the spectral curve: if we look in detail at the jumps in $\omega(\gamma', i)$ 
induced by $2d-4d$ wall-crossing across $\gamma'$ \cite{Gaiotto:2011tf}, we see that the periods of $d\omega(\gamma', i)$ around turning points are zero. 
Then $\partial_{\bar z} -\I d\theta_i$ is the $U(1)$ connection which parameterizes the Higgs bundle, and $\theta_{\gamma}$ coordinates on the fibre.

Notice that there is a certain degree of ambiguity in picking the square roots of the canonical bundles of the base curve 
and the spectral curve. A clean, if unfamiliar, way to eliminate the ambiguity is to take the two square roots as 
twisted line bundles rather than line bundles \cite{2007arXiv0710.5939F}. Then the coordinates $\theta_{\gamma}$ 
are holonomies of a twisted $U(1)$ bundle on the spectral curve, and canonical coordinates on the moduli space 
of Hitchin's equations for a twisted bundle. 

This twisted perspective is not strictly necessary: one can work with normal bundles, making extra non-canonical choices which 
propagate in the form of sign choices in many places, in particular in the choice of quadratic refinements $\sigma(\gamma)$. 
On the other hand, the degeneracies $\mu$ and $\omega$ have been computed in the twisted formalism \cite{Gaiotto:2011tf}, 
where they have canonical, natural signs. 

As the choice of square root of the canonical bundle is needed in order to define the Hitchin zero section
and the oper manifold, the twisted formalism has the added benefit of making these completely canonical. 
 
\subsection{The section of Hitchin fibration}
It is natural to set the $\theta_\gamma$ (and $d \theta_i$) to zero in the TBA equations, i.e. look at the special section of Hitchin's fibration 
associated to a trivial degree-zero line bundle on the spectral curve. 
The solutions of the TBA equations acquire an extra symmetry 
$X_{-\gamma}(- \zeta) = X_\gamma(\zeta)$ which removes the $\rho_i$ corrections as well. 

The $g_i^+$ matrix can then be given in detail. The $k$-th element of $g_i^+$ is 
\begin{equation}
(g_i^+)^k = \prod_{j \neq i} (x_i - x_j)^{-1/2} x_i^k 
\end{equation}
Notice that the determinant of $(g_i^+)^k$ is $1$. 

This form of $g_i^+$ corresponds to a very specific gauge choice. The zero section of the Hitchin fibration can be 
{\it defined} by requiring the existence of a non-trivial line-subbundle which generates the whole bundle when acted upon 
by $\Phi$. The first element of $g_i^+$ defines such a sub-bundle, and the other elements are produced by the action of powers of $\Phi$. 
Thus $\Phi$ takes the form 
\begin{equation}
\Phi = \begin{pmatrix} 0 & 0 &\cdots & 0 & \phi_K \cr 1 & 0 &\cdots & 0 & \phi_{K-1} \cr \cdots & \cdots & \cdots &\cdots& \cdots \cr 0 & 0 & \cdots & 1 & 0 \end{pmatrix}
\end{equation}

The integral equations simplify a bit because of the symmetry $X_{-\gamma}(- \zeta) = X_\gamma(\zeta)$ and the condition $d\theta=0$.
The contribution from opposite BPS rays in the integral equations for the  $x_{\gamma_{ij'}}$ can be combined together as before, 
to give an integral kernel which decays at small and large $\zeta'$. 
The integral equations for the $g_i$ do not change: the symmetry relates $g_{-i}(-\zeta)$ and $g_i(\zeta)$, and thus cannot be used to re-group terms. 

\subsection{The conformal limit}
The conformal limit is now obvious. We get 
\begin{equation} \label{eq:int-1-conf}
x_{\gamma_i}(\epsilon) := Z_{\gamma_i}/\epsilon +
\sum_{\gamma'>0} \omega(\gamma', \gamma_i) \frac{1}{\pi i}
\int_{\ell_{\gamma'}}d \epsilon' \frac{\epsilon}{(\epsilon')^2- \epsilon^2}
\log(1 - X_{\gamma'}(\epsilon')).
\end{equation}
and 
\begin{equation} \label{eq:int-2-conf}
g_k(\epsilon) = g_k^+ + \sum_{\ell \neq k, \gamma_{\ell k}} \mu(\gamma_{\ell k})
\frac{\epsilon}{2 \pi \I} \int_{\ell_{\gamma_{\ell k }}} \frac{\de \epsilon'}{\epsilon'}
\frac{1}{\epsilon' - \epsilon} g_\ell (\epsilon') x_{\gamma_{\ell k}}(\epsilon'),
\end{equation}
and similarly for $g_{-i}$. 

Now, we get to the crucial observation. The integral equations build sections of a flat connection 
\begin{equation}
\CA = \sum_i  g_i a_i g_{-i} + g_i dg_{-i} 
\end{equation}
with a rather specific structure. The first element of $g_i^+$, and thus $g_i$, will define a line sub-bundle 
as long as the integral equations do not require a further gauge transformation for the first element of $g_i$
to be well-defined at turning points. This can be verified by a local analysis, which will be our first example in the next section.

The oper manifold can be defined by the existence of such a line sub-bundle, with the property that acting with powers the 
$D_z$ component of the connection generates the whole bundle. The latter condition is an open condition, and 
$D_z$ is dominated by $\Phi$ for small $\epsilon$, thus we expect the solutions of the integral equations will lie in the oper manifold for sufficiently 
small $\epsilon$. This motivates our conjecture that $\CL_\epsilon$ is the oper manifold in $\CM_\epsilon$. 

Notice that we can restrict the integral equations to the first element $f_i$ of $g_i$, which correspond to 
writing the oper as a degree $K$ differential operator, acting on sections of the $(-K/2)$-th power of the canonical bundle on $C$.
 
\subsection{Examples}
In this section we will restrict ourselves to $A_1$ examples with spectral curve
\begin{equation}
x^2 = \phi_2(z).
\end{equation} 
The integral equations involve two functions 
$f_1$ and $f_2$. The integral equation has a symmetry which implies that $f_2(-\epsilon) = \I f_1(\epsilon)\equiv \I \psi(\epsilon)$. 
Thus we can reduce ourselves to a single integral equation 
\begin{equation} \label{eq:int-22}
f(\epsilon) = \frac{1}{\sqrt{2 x}} - \sum_{\gamma_{+-}} \mu(-\gamma_{+-})
\frac{\epsilon}{2 \pi} \int_{\ell_{\gamma_{+-}}} \frac{\de \epsilon'}{\epsilon'}
\frac{1}{\epsilon' + \epsilon} f(\epsilon') x_{\gamma_{+-}}(\epsilon').
\end{equation}
\subsubsection{Airy}
The local behaviour of the $g_i$ integral equations near a turning point is captured by the spectral curve 
\begin{equation}
x^2 = z
\end{equation}
There are no $\omega$ BPS degeneracies, and a single non-zero $\mu$. 
The integral of $\lambda$ along the path $p$ from $z$ to the turning point at the origin and back on the opposite sheet 
is 
\begin{equation}
Z_p=-\frac{4}{3} z^{\frac{3}{2}}
\end{equation}
and the integral equation becomes 
\begin{equation} \label{eq:int-22-Airy}
f(\epsilon) = \frac{1}{\sqrt{2 x}} -\frac{\epsilon}{2 \pi} \int_{\ell_{p}} \frac{\de \epsilon'}{\epsilon'}
\frac{1}{\epsilon' + \epsilon} f(\epsilon') e^{-\frac{4}{3 \epsilon'} z^{\frac{3}{2}}}.
\end{equation}

The solution converges rapidly to 
\begin{equation}
f(\epsilon) = \frac{\sqrt{2 \pi}}{\epsilon^{1/6}} e^{\frac{2}{3 \epsilon} z^{\frac{3}{2}}} \mathrm{Ai}(\frac{z}{\epsilon^{2/3}})
\end{equation}
which gives us the flat section $\frac{\sqrt{2 \pi}}{\epsilon^{1/6}} \mathrm{Ai}(\frac{z}{\epsilon^{2/3}})$
of the Airy oper
\begin{equation}
-\epsilon^2 \partial_z^2 + z 
\end{equation}

\section{An improved parameterization: solving inverse scattering problems}
As it stands, the set of integral equations we described produces a parameterization of the oper manifold and the flat sections for the opers
which is somewhat convoluted: everything is expressed in terms of the periods $Z_\gamma$. 

A natural problem one may consider is to identify an oper, say a Schr\"odinger operator, with prescribed scattering data/ cross-ratios, and find the corresponding flat sections. 
The current form of the integral equations is not quite convenient for the purpose. We can easily amend that, using a trick from \cite{Alday:2010ku}.
Start from the integral equation in the conformal limit \ref{eq:int-old} specialized to 
$\epsilon =1$, 
\begin{equation} \label{eq:int-old1}
\log X_{\gamma}(1) = Z_\gamma +  \sum_{\gamma'>0} \omega(\gamma', \gamma) \frac{1}{\pi \I} \int_{\ell_{\gamma'}} \frac{\de \epsilon'}{(\epsilon')^2 - 1} \log(1 - \sigma(\gamma')X_{\gamma'}(\epsilon')).
\end{equation}
solve for $Z_\gamma$ and plug back into \ref{eq:int-old}:
\begin{equation} \label{eq:int-old-new}
\log X_{\gamma}(\epsilon) = \frac{\log X_{\gamma}(1)}{\epsilon} +  \sum_{\gamma'>0} \omega(\gamma', \gamma) \frac{\epsilon-\epsilon^{-1}}{\pi \I} \int_{\ell_{\gamma'}} \frac{\de \epsilon'}{(\epsilon')^2 - (\epsilon)^2} \frac{(\epsilon')^2}{(\epsilon')^2-1} \log(1 - \sigma(\gamma')X_{\gamma'}(\epsilon')).
\end{equation}

For complete gauge theories, such as $A_1$ examples associated to the Schr\"odinger operators, there should be no problem associated with solving for $Z_\gamma$:
by varying all the parameters in the quadratic differential, including the position of punctures, one can reach generic values of $Z_\gamma$. 
Thus one can solve the above integral equation to find the choice of periods $Z_\gamma$ which give specific scattering data/crossratios $X_{\gamma}(1)$ at $\epsilon=1$. 

\section*{Acknowledgements}
The research of DG was supported by the Perimeter Institute for Theoretical Physics. Research at Perimeter Institute is supported by the Government of Canada through Industry Canada and by the Province of Ontario through the Ministry of Economic Development and Innovation.

\bibliographystyle{JHEP}

\bibliography{swn-paper}

\end{document}